\documentclass[acmlarge]{acmart}
\makeatletter
\newcommand{\myconfshort}{\acmConference@shortname}
\newcommand{\myconffull}{\acmConference@name}
\newcommand{\myconfdate}{\acmConference@date}
\newcommand{\myconfloc}{\acmConference@venue}
\AtBeginDocument{
  \fancypagestyle{firstpagestyle}{
    \fancyhead{}%
    \fancyfoot[C]{}%
  }
  \fancyhf{}
  \fancyhead[LO]{\@headfootfont\shorttitle}%
  \fancyhead[RE]{\@headfootfont\@shortauthors}%
  \fancyhead[LE]{\@headfootfont\footnotesize \myconfshort, \myconfdate, \myconfloc}%
  \fancyhead[RO]{\@headfootfont\footnotesize \myconfshort, \myconfdate, \myconfloc}%
  \fancyfoot[C]{}%
}
\makeatother
\acmBooktitle{\conffull\@ (\confshort), \confdate, \confloc}

\AtBeginDocument{%
  }

\copyrightyear{2026}
\acmYear{2026}
\setcopyright{cc}
\setcctype{by}
\acmConference[FAccT '26]{The 2026 ACM Conference on Fairness, Accountability, and Transparency}{June 25--28, 2026}{Montreal, QC, Canada}
\acmBooktitle{The 2026 ACM Conference on Fairness, Accountability, and Transparency (FAccT '26), June 25--28, 2026, Montreal, QC, Canada}
\acmDOI{10.1145/3805689.3812380}
\acmISBN{979-8-4007-2596-8/2026/06}

\usepackage{multirow}
\usepackage{tikz}
\usetikzlibrary{arrows.meta, positioning}

\begin{document}

\title[Does Algorithmic Uncertainty Sway Human Experts?]{Does Algorithmic Uncertainty Sway Human Experts? Evidence from a Field Experiment in Selective College Admissions}

\author{Hansol Lee}
\affiliation{%
  \institution{Stanford University}
  \city{Stanford}
  \state{CA}
  \country{USA}}
\email{hansol@stanford.edu}

\author{AJ Alvero}
\affiliation{%
  \institution{Cornell University}
  \city{Ithaca, NY}
  \country{USA}}
\email{ajalvero@cornell.edu}

\author{Ren\'e F. Kizilcec}
\affiliation{%
  \institution{Cornell University}
  \city{Ithaca, NY}
  \country{USA}}
\email{kizilcec@cornell.edu}

\author{Thorsten Joachims}
\affiliation{%
  \institution{Cornell University}
  \city{Ithaca, NY}
  \country{USA}}
\email{tj@cs.cornell.edu}



\begin{abstract}
Algorithmic predictions are inherently uncertain: even models with similar aggregate accuracy can produce different predictions for the same individual, raising concerns that high-stakes decisions may become sensitive to arbitrary modeling choices. In this paper, we define \emph{algorithmic sensitivity} as the extent to which arbitrary modeling choices propagate into human decisions: how much a decision outcome shifts when a more favorable versus less favorable algorithmic prediction is presented to the decision-maker for the same individual. We estimate this in a randomized field experiment ($n=19{,}545$) embedded in a selective U.S. college admissions cycle, in which admissions officers reviewed each application alongside an algorithmic score while we randomly varied whether the score came from one of two similarly accurate prediction models. Although the two models performed similarly in aggregate, they frequently assigned different scores to the same applicant, creating exogenous variation in the score shown. Surprisingly, we find little evidence of algorithmic sensitivity: presenting a more favorable score does not meaningfully increase an applicant's probability of admission on average, even when the models disagree substantially. These findings suggest that, in this expert, high-stakes setting, human decision-making is largely invariant to arbitrary variation in algorithmic predictions, underscoring the role of professional discretion and institutional context in mediating the downstream effects of algorithmic uncertainty.
\end{abstract}




\maketitle

\section{Introduction}

Predictive models are increasingly used to inform high-stakes social decisions, from criminal justice and lending to medical diagnosis, hiring, and college admissions \cite{kleinbergHumanDecisionsMachine2017, stevenson2024algorithmic}. The case for using them rests on the complementary strengths of machines and humans. Models can synthesize structured information at scale and with a level of consistency that human decision-makers, working under time pressure and subject to noisy judgment, often cannot achieve \cite{meehl1954clinical, kahnemanNoiseHowOvercome2016}. Human decision-makers, in turn, bring contextual knowledge, professional judgment, and the capacity to integrate information that is not encoded in training data. The promise of human-algorithm collaboration is that the combination can perform better than either alone \cite{steyvers2022bayesian, bansal2021does}.

At the same time, the algorithmic prediction assigned to an individual is not a fixed property of that individual. Similarly accurate models can produce different predictions for the same individual \cite{black2022model, marx2020predictive}, predictions can shift as models are retrained on new data \cite{gama2014survey, quinonero2022dataset}, and choices made in feature engineering, calibration, and model selection can each move a prediction up or down \cite{d2022underspecification}. Encoded biases in training data introduce further variation \cite{barocas2016big}. The algorithmic prediction a human decision-maker sees is therefore shaped in part by aspects of the modeling process that often have little to do with the underlying individual. This raises a concern that has motivated a substantial literature on algorithmic fairness and accountability: whether decisions made downstream of such predictions end up depending on arbitrary modeling choices rather than substantively meaningful differences between individuals \cite{cooper2024arbitrariness, wang2024against}.

In response to these concerns, a dominant decision-making pattern has emerged in which the algorithm is positioned not as a decision-maker but as a decision aid, with trained human experts retaining final authority over consequential decisions. College admissions is one such setting. Virginia Tech, for instance, recently announced a new undergraduate admissions process that pairs human essay readers with an AI-supported scoring system developed by university researchers \cite{virginia-tech}, with final decisions made exclusively by trained admissions professionals. Such arrangements are widespread: industry surveys suggest roughly half of higher-education admissions offices were already using AI tools by 2023 \cite{intelligent2023ai}. The justifying logic is often that the human-in-the-loop will serve as a corrective force, absorbing the imperfections of the algorithm rather than passing them through to outcomes.

Whether this safeguard actually works depends on a specific empirical question. When a human decision-maker sees a more favorable rather than a less favorable score for the same individual, is their decision swayed by which particular score they happen to see? If decisions track the score, then arbitrary aspects of the modeling process propagate into outcomes even when humans nominally retain authority. If decisions are largely invariant to the particular score shown, then expert judgment is performing much of the corrective role that human-in-the-loop arrangements presuppose. We call this property \textbf{algorithmic sensitivity}: the extent to which arbitrary modeling choices propagate into consequential human decisions, irrespective of whether the underlying decisions are correct.

Existing evidence on this question is mixed and largely comes from laboratory experiments or observational field settings, where decision-makers and algorithmic input are not randomly paired. The most direct test would be a field experiment in which decision-makers are systematically shown a more favorable score for some individuals and a less favorable score for others, but such a design would be ethically impermissible: it would knowingly disadvantage some individuals by presenting decision-makers with information that is, by construction, less favorable than what those individuals would otherwise have received. Our approach engages this challenge by exploiting \emph{predictive multiplicity}, the phenomenon that similarly accurate models can produce different predictions for the same individual \cite{black2022model, marx2020predictive}. When two such models disagree, the score a decision-maker happens to see is more or less favorable purely as a function of which model was deployed. Randomizing which of two such models' scores is presented therefore induces exogenous variation in the favorability of the displayed score without assigning any individual a score from a model known to be inferior. Predictive multiplicity is one of several sources of model-contingent variation in algorithmic scores, and findings about sensitivity to this form of variation have implications for the broader concern.

We apply this design in a randomized field experiment embedded in a selective U.S.\ college admissions process ($n=19{,}545$ applications) during the 2022 admissions cycle, in close collaboration with the university's admissions leadership \cite{lee2024algorithms, lee2023evaluating}. We developed two predictive models using the same features and modeling pipeline but trained on slightly different historical data, reflecting realistic variation in modeling choices. The two models achieve similar aggregate performance but assign different score deciles to over 70\% of applicants. For each application, we randomized which model's score was presented, allowing us to identify the causal effect of presenting a more favorable versus less favorable algorithmic prediction for the same applicant.

Despite the substantial individual-level disagreement, we find little evidence of algorithmic sensitivity. Presenting a more favorable score does not meaningfully increase an applicant's probability of admission on average, even when the models disagree substantially. Across specifications, admission outcomes are largely invariant to which of the two disagreeing algorithmic scores is shown. These findings suggest that, in this expert, high-stakes decision-making setting, model-contingent variation in algorithmic scores does not translate directly into variation in admission outcomes. Rather than responding to which particular algorithmic score they observe, admissions officers appear to integrate algorithmic scores as one input among many within a holistic evaluation process.

This paper makes three contributions. Conceptually, we propose \emph{algorithmic sensitivity} as a framework for studying whether arbitrary variation in algorithmic input propagates into consequential human decisions, applicable to domains where no canonical ground truth exists against which to evaluate decisions. Methodologically, we show that predictive multiplicity can serve as a source of exogenous variation for identifying algorithmic sensitivity in real institutional settings, opening up causal analysis where direct manipulation of algorithmic input would be ethically impermissible. Empirically, we apply this design in a randomized field experiment in selective college admissions and find that reviewers' decisions are largely invariant to model-contingent variation in algorithmic scores, suggesting that the downstream consequences of algorithmic uncertainty depend on the sociotechnical context in which the algorithm is embedded, not just on properties of the model.

\section{Related Work}
\label{sec:related}

Our study sits at the intersection of three concerns that have been pursued largely in parallel: a literature on how human decision-makers respond to algorithmic input, a literature on predictive uncertainty and the arbitrariness of algorithmic outputs, and a smaller body of recent work that studies algorithm-assisted decision-making in real institutional contexts. We outline what each contributes and identify the gaps that motivate our framework and design.

\subsection{Human Reliance on Algorithmic Advice}

Early experimental work on human responses to algorithmic advice documented two opposing tendencies: \emph{algorithm aversion}, in which decision-makers discount algorithmic forecasts after observing them err, even when the algorithm continues to outperform the available human alternative \cite{dietvorst2015algorithm}, and \emph{algorithm appreciation}, in which lay decision-makers weight algorithmic advice more heavily than equivalent human advice \cite{logg2019algorithm}. A related literature on \emph{automation bias} examines settings in which decision-makers over-defer to algorithmic outputs, failing to detect errors even when they have access to information that would allow them to do so \cite{goddard2012automation, passi2022overreliance}. Even nominally human-in-the-loop designs can degrade decision quality when human oversight is mechanical rather than discretionary: \citet{sele2024putting} show that human monitors are less likely to intervene on the least accurate recommendations, raising concerns about whether human oversight achieves the corrective function that often justifies its inclusion \cite{green2022flaws}.

When the decision-maker is an expert working in a high-stakes domain, the question takes on a sharper form: does professional judgment work as the intended corrective to algorithmic limitations? Recent work in real institutional settings suggests that it can. \citet{cheng2022child} analyzed call-screen workers at the Allegheny County Department of Human Services using an algorithmic family screening tool, and found that workers exercising holistic judgment substantially reduced racial disparities relative to the algorithm alone, narrowing the gap in screen-in rates between Black and white children from 20\% to 9\%. \citet{de2020case} document a related pattern in child welfare hotline screening: workers were less likely to adhere to the algorithm's recommendation when the displayed score was an incorrect estimate of risk, even when overriding required supervisory approval. \citet{schechtman2025discretion} study an algorithm-assisted advising program at Georgia State University with a randomized controlled trial design, and estimate that roughly two-thirds of advisor interventions were plausibly targeted using non-algorithmic context, with advising style itself shaping student outcomes.

But expert discretion does not always serve a corrective role. \citet{albright2019if} examines a Kentucky bail reform that set a recommended default based on algorithmic risk scores, and finds that the recommended default was disproportionately overridden in favor of harsher bond conditions for Black defendants relative to comparable white defendants. The result was that judicial discretion amplified, rather than attenuated, racial disparities in pretrial detention. \citet{green2021algorithmic} provide complementary lab-based evidence that algorithmic risk assessments can systematically alter how decision-makers weight risk relative to competing considerations, in ways that can undermine policy goals even when prediction accuracy improves.

Read together, this literature establishes that the relationship between algorithmic input and expert decisions is real, consequential, and highly contextual. It does not, on its own, tell us whether human reviewers will buffer or transmit arbitrary variation in the algorithmic input itself. A second feature of this literature also limits how directly it bears on our question: much of it evaluates human reliance against a known ground truth, asking whether deference to the algorithm improves decision accuracy or fairness with respect to an observable outcome \cite{guo2024decision, schemmer2023appropriate, donahue2022human}. This is less applicable to college admissions and other social decisions that weigh multiple incommensurable criteria and lack a single objectively correct answer \cite{bruch2017decision, stevens2009creating, michel2019graduate}. Our framework of algorithmic sensitivity takes up this gap: by asking only whether different scores lead to different decisions for the same individual, it decouples the question of how algorithmic input propagates through expert judgment from the question of whether the algorithm or the human is right.

\subsection{Predictive Multiplicity and Algorithmic Arbitrariness}

A second literature speaks to the property of algorithmic systems that motivates our methodological design. \citet{breiman2001statistical}'s early observation of the \emph{Rashomon Effect}, that many distinct models can achieve similar aggregate performance on the same task, has been formalized in recent years as \emph{predictive multiplicity} \cite{marx2020predictive, black2022model} and \emph{underspecification} \cite{d2022underspecification}: equally accurate models can yield meaningfully different predictions for the same individual. A growing normative literature treats this contingency as a problem for the legitimacy of algorithmic decision-making. When outcomes depend on which of several equally defensible models is deployed, decisions become difficult to justify on substantive grounds \cite{cooper2024arbitrariness}, individuals bear the cost of these arbitrary choices \cite{ganesh2024cost}, and these concerns extend into a broader argument against predictive optimization as a foundation for consequential decisions \cite{wang2024against}.

What is largely absent from this literature is empirical evidence on the question that determines whether the concern about model-level arbitrariness translates into a concern about decision-level arbitrariness in deployed systems. The literature documents that model outputs can differ for the same individual under defensible modeling choices; it generally does not test whether such differences propagate through the human decision-makers who actually issue consequential decisions in the institutional settings where these models are deployed. Our methodological move turns predictive multiplicity from a property to be characterized into a tool: by randomly assigning which of two equally accurate models' scores a reviewer sees, we obtain exogenous variation in the favorability of the algorithmic input and can directly identify whether that variation propagates into outcomes.

\subsection{Field-Based Studies of Algorithm-Assisted Decision-Making}

A smaller body of work studies algorithm-assisted decision-making in its actual institutional context. This work is closest in spirit to ours, both in setting and in methodological ambition, and it shapes how we interpret what we find.

\citet{stevenson2024algorithmic} study the introduction of algorithmic risk assessments for felony sentencing in Virginia. Their finding is nuanced: judges' decisions were influenced by the risk scores, with higher scores leading to longer sentences and lower scores to shorter ones, but judicial discretion mediated the tool's impact in systematic ways. Notably, judges granted leniency to young defendants despite their high risk scores, and over time used the scores less. The introduction of the tool did not produce the public safety gains its designers had anticipated, illustrating how expert discretion can substantially reshape the practical effect of algorithmic input even when input does enter decisions. \citet{imai2023experimental} develop a statistical framework for experimentally evaluating algorithm-assisted human decision-making and apply it to a randomized controlled trial of the pretrial Public Safety Assessment, finding that providing the PSA to judges had little overall impact on judges' decisions or subsequent arrestee behavior. \citet{brayne2021technologies} provide ethnographic evidence from a large urban police department and a midsized criminal court, showing that algorithmic tools meet substantial professional resistance and that, where they are absorbed, they tend to relocate discretion to less visible and less accountable areas of organizational practice rather than displacing it entirely.

Across these studies, a common pattern emerges: the downstream consequences of algorithmic input in real institutional settings are often more muted, and more mediated by professional judgment, than analyses of model behavior in isolation would predict. Our empirical finding sits comfortably within this pattern, and the design we develop extends this body of work in two ways. Conceptually, prior field-based work largely retains the reliance framework's implicit appeal to ground truth, which constrains its portability to domains like admissions; our framework of algorithmic sensitivity is designed to be usable where that appeal is not available. Methodologically, prior studies vary the presence or salience of an algorithmic input, while we vary which of two equally defensible algorithmic inputs is shown, holding the institutional context, the reviewer, and the application constant. This lets us isolate the effect of the input's specific value from the effect of the input's presence, and to do so without imposing a manipulation that disadvantages applicants relative to what they would otherwise have received.

\section{Context: Algorithm-Assisted College Admissions at a Selective U.S.\ University}
\label{sec:context}

Selective college admissions is a useful setting for studying how human experts respond to algorithmic input. Admissions decisions are not reducible to ranking applicants by a single notion of merit; they involve constructing a class that satisfies multiple, often competing institutional goals \cite{stevens2009creating}. Admissions officers therefore engage in holistic review, synthesizing quantitative indicators with qualitative assessments of context, potential, and institutional fit \cite{michel2019graduate}. As a result, admissions decisions are inherently judgment-based and normatively contested rather than objectively correct \cite{bruch2017decision}. In this setting, the question of whether algorithmic input propagates into decisions can be asked directly, without recourse to a notion of decision correctness.

Our study is situated at a highly selective U.S.\ university during the 2021--2022 admissions cycle, following the widespread adoption of test-optional policies in response to the COVID-19 pandemic. These policies substantially reduced the availability of standardized test scores \cite{bennett2022untested}, historically a central quantitative input into admissions decisions. At the same time, application volumes rose, placing additional time and attention constraints on admissions officers. To support holistic review under these conditions, the institution introduced predictive algorithms trained on historical admissions data. The goal was not to automate admissions decisions but to provide a test-free signal that could help reviewers allocate limited attention and ensure that strong applicants were not inadvertently overlooked. Details on the design of these algorithms are reported in prior work \cite{lee2023evaluating, lee2024algorithms}.

The algorithms produce predicted probabilities of admission based on information available at the time of application, including academic records and other observed characteristics. For operational use, these probabilities are discretized into deciles (1--10, with 10 indicating the highest predicted probability) and displayed to admissions officers alongside each applicant's materials. The scores are intended as a coarse, directional signal that reviewers can use to prioritize higher-scoring applications for earlier review or to flag applications that might otherwise receive limited attention. They are not used as thresholds at any stage of the process: no applicant is automatically advanced, eliminated, or routed based on the score alone. Final admissions decisions remain with trained admissions professionals, who exercise holistic judgment based on essays, recommendation letters, contextual background, and institutional priorities not captured by the model.

Algorithmic scoring in this context raises ethical concerns specific to admissions that lie beyond the scope of the empirical question we study. Models trained on historical admissions data risk encoding patterns shaped by structural disadvantages \cite{barocas2016big, gandara2024inside}, and prior work at this institution shows that removing race data from applicant ranking algorithms reduces the diversity of the top-ranked pool without meaningfully improving academic merit, while increasing arbitrariness in outcomes for most applicants \cite{lee2024algorithms}. Our findings on algorithmic sensitivity do not resolve these concerns or speak to whether the deployed models are themselves fair. The Ethical Considerations Statement examines the design of this study in greater depth.

\section{Methods}

We define algorithmic sensitivity as a population-level causal estimand and develop the experimental design that permits its identification in a field setting.

\subsection{Defining algorithmic sensitivity}
\label{sec:ar}

We study whether decision outcomes for the same individual depend on which algorithmic prediction is presented when multiple plausible predictions exist. We define \textit{algorithmic sensitivity (AS)} as the population-level causal effect of presenting a more favorable algorithmic prediction rather than a less favorable one on downstream decisions, holding fixed the individual.

Let $A$ denote the space of algorithmic predictions that may be presented to a decision-maker. For any $a\in A$, let $Y_i(a)\in\{0,1\}$ denote the potential decision outcome for individual $i$ if prediction $a$ were presented, where $Y_i(a)=1$ indicates a favorable decision outcome (e.g., admission).

Let $\succ$ denote an ordering over $A$ such that $a^H \succ a^L$ means $a^H$ is more favorable than $a^L$. For any ordered pair $(a^H,a^L)\in A\times A$ with $a^H \succ a^L$, define \emph{algorithmic sensitivity} as
\begin{equation}
\text{AS}(a^H,a^L)
\;:=\;
\mathbb{E}\!\left[ Y_i(a^H) - Y_i(a^L) \right].
\label{eq:ar-pair}
\end{equation}

Under this definition, algorithmic sensitivity has the following behavioral interpretation:
\begin{itemize}
    \item $\text{AS}(a^H,a^L) \approx 0$: showing $a^H$ rather than $a^L$ does not materially change decisions.
    \item $\text{AS}(a^H,a^L) > 0$: showing $a^H$ rather than $a^L$ \textit{increases} the likelihood of a favorable decision; larger values indicate stronger sensitivity to the algorithmic signal.
    \item $\text{AS}(a^H,a^L) < 0$: showing $a^H$ rather than $a^L$ \textit{decreases} the likelihood of a favorable decision; larger negative values indicate stronger counteracting or aversive responses to the algorithmic signal.
\end{itemize}

This definition does not assume the existence of a ground-truth or objectively correct decision, nor does it require that either the human decision or the algorithmic prediction be accurate. Algorithmic sensitivity is defined purely in terms of how observed decisions respond to changes in the algorithmic input.

For example, if across individuals decisions are favorable 40\% of the time when $a^H$ is shown and 30\% of the time when $a^L$ is shown, then $\text{AS}(a^H,a^L)=0.10$: presenting $a^H$ rather than $a^L$ increases the probability of a favorable decision by 10 percentage points on average. In settings characterized by predictive multiplicity, algorithmic sensitivity captures whether model-contingent differences in algorithmic scores translate into differences in real outcomes for the same individual.

\subsection{Experiment Design}
\label{sec:experiment-design}

Our field experiment was conducted during the 2022 Regular Decision admissions cycle at a highly selective U.S.\ university with $n = 19{,}545$ applications, under a research protocol approved by an Institutional Review Board (IRB protocol number \#2001009365) and in close collaboration with senior admissions leadership at the participating institution. The design exploits \emph{predictive multiplicity}, the fact that multiple predictive models can achieve similar aggregate performance while producing different scores for the same individual \cite{black2022model, marx2020predictive}. Rather than manipulating the favorability or quality of algorithmic advice, our design exploits naturally occurring, model-contingent variation among equally plausible predictions. This generates exogenous variation in the algorithmic score observed by decision-makers without degrading model accuracy or assigning any applicant a score from a model known to be inferior. Figure~\ref{fig:dag} in Appendix~\ref{appendix:graph} illustrates the causal structure of the experimental design and highlights the specific edge under study.

\subsubsection{Model development}

We developed two predictive models, Model~1 and Model~2, using the same modeling pipeline but slightly different training data. Both models are gradient boosted decision tree classifiers trained to predict the probability of admission, following institutional practice and prior work in this setting \cite{lee2023evaluating, lee2024algorithms}.\footnote{Predicting admission probability grounds the signal in the institution's own historical decisions, which is appropriate for a tool designed to support attention allocation within that same process. Alternative targets such as post-admission outcomes (e.g., GPA) were unavailable at application time and would not reflect the multi-criteria nature of admissions decisions, which involve constructing a class that satisfies multiple institutional goals beyond academic performance alone \cite{stevens2009creating}.}

The data used to train the two models comes from the university's 2019--20 and 2020--21 admissions cycles, including all information submitted via the Common Application: academic records (high school GPA, class rank, courses taken, AP/IB scores, and optional SAT subject scores) and personal information (essays, extracurricular activities, honors and awards, intended major, legacy status, and demographic indicators including gender, race and ethnicity, and first-generation status). Recommendation letters and school-specific counselor reports were not available. SAT/ACT and TOEFL/IELTS scores were explicitly excluded to simulate the test-optional environment of the experiment.

The two models differ only in their training data, reflecting realistic variation in modeling choices one would encounter in practice. Model~1 was trained on applicants from the 2020--21 Regular Decision cycle only; Model~2 was trained on both the 2019--20 and 2020--21 cycles. Each model outputs a predicted probability of admission, discretized into deciles (1--10) before being displayed to reviewers, consistent with institutional practice.

\begin{table}[ht]
\centering
\caption{Out-of-sample model validation performance.}
\label{tab:model-performance}
\begin{tabular}{lcccc}
\toprule
& \multicolumn{2}{c}{2020--21 RD} & \multicolumn{2}{c}{2021--22 ED} \\
\cmidrule(lr){2-3} \cmidrule(lr){4-5}
& AUROC & Avg.\ Precision & AUROC & Avg.\ Precision \\
\midrule
Model~1 & 86.9 & 32.5 & 80.9 & 52.1 \\
Model~2 & 87.8 & 34.5 & 82.0 & 54.9 \\
\bottomrule
\end{tabular}
\end{table}

As reported in Table~\ref{tab:model-performance}, the two models exhibit similar aggregate predictive performance on out-of-sample validation data. Model~2 performs slightly better on average in terms of AUROC and Average Precision, reflecting its larger training set. However, the inclusion of older data also increases the risk of distributional shift \citep{quinonero2022dataset}, making it uncertain whether this advantage would persist in the upcoming cycle---part of what motivates treating both models as equally defensible. Figure~\ref{fig:2022} also shows the two models are similarly calibrated with respect to observed decisions, supporting the comparability of their score scales. While aggregate performance is similar, the models show substantial individual-level disagreement: \textbf{the two models assign different score deciles to 73.2\% of applicants} in the experimental sample, providing the variation we exploit to study algorithmic sensitivity (Section~\ref{sec:results-disagreement}).

\subsubsection{Randomized Algorithmic Score Presentation}

For each applicant $i$, both model scores are computed offline prior to human review, denoted $(S_{i1}, S_{i2}) \in \{1,\ldots,10\}^2$. Exactly one of these scores is shown to admissions officers, with assignment $W_i \sim \mathrm{Bernoulli}(0.5)$, where $W_i = 0$ indicates that the Model~1 score is presented and $W_i = 1$ that the Model~2 score is presented. All other aspects of the applications and the decision-making processes are held fixed. Admissions officers were blind to which model produced the scores they were presented with. 


Randomization yielded 9,765 applications assigned to the Model~1 condition and 9,780 to the Model~2 condition. The outcome of interest, $Y_i \in \{0,1\}$, indicates whether applicant $i$ was ultimately admitted; the overall admission rate in the analytic sample is 5.44\%. Applicant characteristics are well balanced across experimental arms (Table~\ref{tab:covariate-balance}), consistent with successful randomization.

\begin{table}[ht]
\centering
\caption{Proportion of applicants by demographic group across the two experimental conditions.}
\label{tab:covariate-balance}
\begin{tabular}{lcccc}
\toprule
Condition & First-generation & International & URM & Female \\
\midrule
Model~1 presented & 0.171 & 0.269 & 0.144 & 0.298 \\
Model~2 presented & 0.165 & 0.262 & 0.147 & 0.310 \\
\bottomrule
\end{tabular}
\end{table}

\subsection{Estimating algorithmic sensitivity}
\label{sec:estimating-ar}

Our experimental design randomizes which model's score is displayed, not whether reviewers see a more favorable versus a less favorable prediction. When the two models disagree for a given applicant, however, the randomized model assignment induces exogenous variation in the favorability of the displayed score. We show how this permits identification and unbiased estimation of algorithmic sensitivity for the subset of prediction pairs that arise from model disagreement.

\subsubsection{Setup and estimand}

We use the notation introduced in Section~\ref{sec:experiment-design}: $(S_{i1}, S_{i2}) \in \{1,\ldots,10\}^2$ are the two model scores for applicant $i$, $W_i \in \{0,1\}$ indicates which is displayed, and $Y_i \in \{0,1\}$ is the realized admission outcome.

Define the disagreement set $\mathcal{D}:=\{i: S_{i1}\neq S_{i2}\}$. For each $i\in\mathcal{D}$, define the more and less favorable realized scores as
\[
H_i := \max(S_{i1},S_{i2}), \qquad
L_i := \min(S_{i1},S_{i2}).
\]

Let $Y_i(H_i)$ and $Y_i(L_i)$ denote the potential admission outcomes if the more favorable versus less favorable score were shown.\footnote{This potential-outcomes notation invokes the Stable Unit Treatment Value Assumption (SUTVA): applicant $i$'s outcome under a given displayed score depends only on the score shown for applicant $i$, not on scores shown for other applicants. In principle, admissions decisions could exhibit interference through class-size constraints. We expect any such spillover to be negligible in practice given the size of the applicant pool ($n=19{,}545$) relative to the admitted class, the absence of within-cycle communication between reviewers about score assignments, and the balanced random assignment across applicants.} We define algorithmic sensitivity in our setting as
\begin{equation}
\text{AS}
:= \mathbb{E}\!\left[Y_i(H_i)-Y_i(L_i)\mid i\in\mathcal{D}\right].
\label{eq:ar}
\end{equation}

Equivalently, $\text{AS} = \mathbb{E}\!\left[\text{AS}(H_i, L_i) \mid i \in \mathcal{D}\right]$ in the notation of equation~(\ref{eq:ar-pair}): the AS estimand is the expectation of the pair-level algorithmic sensitivity over the distribution of disagreement pairs in $\mathcal{D}$.

\subsubsection{Decomposition}

By the law of iterated expectations,
\begin{align}
\text{AS}
&=
\underbrace{\Pr(S_{i1}>S_{i2}\mid i\in\mathcal{D})}_{\omega_{1>2}}
\cdot
\underbrace{\mathbb{E}[Y_i(H_i)-Y_i(L_i)\mid S_{i1}>S_{i2}]}_{\text{AS}_{1>2}}
\nonumber\\
&\quad+
\underbrace{\Pr(S_{i1}<S_{i2}\mid i\in\mathcal{D})}_{\omega_{2>1}}
\cdot
\underbrace{\mathbb{E}[Y_i(H_i)-Y_i(L_i)\mid S_{i1}<S_{i2}]}_{\text{AS}_{2>1}}.
\label{eq:decomp}
\end{align}

Here $\omega_{1>2}$ and $\omega_{2>1}$ are the frequencies of the two disagreement directions, and
$\text{AS}_{1>2}$ and $\text{AS}_{2>1}$ are the corresponding stratum-specific components of algorithmic sensitivity.

\subsubsection{Identification}

Within $\mathcal{D}$, whether the higher score is shown is a deterministic function of $(S_{i1},S_{i2})$ and $W_i$:
\begin{itemize}
    \item If $S_{i1}>S_{i2}$, the higher score is shown if and only if $W_i=0$.
    \item If $S_{i1}<S_{i2}$, the higher score is shown if and only if $W_i=1$.
\end{itemize}

Since $W_i$ is randomized independently of $(S_{i1}, S_{i2})$ and the potential outcomes $\{Y_i(a) : a \in A\}$, the stratum-specific components satisfy
\[
\text{AS}_{1>2}
=
\mathbb{E}[Y_i\mid W_i=0,S_{i1}>S_{i2}]
-
\mathbb{E}[Y_i\mid W_i=1,S_{i1}>S_{i2}],
\]
and
\[
\text{AS}_{2>1}
=
\mathbb{E}[Y_i\mid W_i=1,S_{i1}<S_{i2}]
-
\mathbb{E}[Y_i\mid W_i=0,S_{i1}<S_{i2}].
\]

In our analysis, we therefore estimate $\text{AS}_{1>2}$ and $\text{AS}_{2>1}$ by sample differences in means and combine them using the empirical weights $\widehat{\omega}_{1>2}$ and $\widehat{\omega}_{2>1}$ to obtain
\[
\widehat{\text{AS}}=\widehat{\omega}_{1>2}\widehat{\text{AS}}_{1>2}+\widehat{\omega}_{2>1}\widehat{\text{AS}}_{2>1}.
\]

Each stratum-specific component ($\widehat{\text{AS}}_{1>2}$ and $\widehat{\text{AS}}_{2>1}$) is computed as the difference in sample means of $Y_i$ across the two values of $W_i$ within the stratum, equivalent to the coefficient on $W_i$ in an unadjusted linear probability model $Y_i = \alpha + \beta W_i + \epsilon_i$ fit within the stratum. Standard errors reported alongside our estimates are HC3 robust standard errors from this regression.

\section{Results}

Our analysis proceeds in three steps. First, we establish that the algorithmic scores shown to admissions officers are meaningful inputs to admissions decisions: higher scores are associated with higher admission probabilities, and this relationship is similar between the two models. Second, we characterize the nature of the algorithmic variation introduced by the experiment, documenting how often and in what ways the two models disagree in their predictions of the same applicant. Finally, leveraging this randomized variation in score presentation, we estimate algorithmic sensitivity: whether admission decisions are systematically influenced by which of two disagreeing but equally plausible algorithmic scores is shown. Together, these analyses allow us to assess whether individual-level algorithmic instability propagates into downstream admission decisions made by admissions officers.

\subsection{How do algorithmic scores relate to admission decisions?}

Before presenting our estimates of algorithmic sensitivity, we first examine how the algorithmic scores used in the experiment relate to downstream admission outcomes. Understanding this relationship is important for two reasons. Substantively, it establishes whether the scores are meaningful inputs into the decision process. Methodologically, it clarifies whether the two models operate on comparable score scales, which matters for interpreting the experimental manipulation.

\begin{figure}[h]
\centering
\includegraphics[width=0.65\columnwidth]{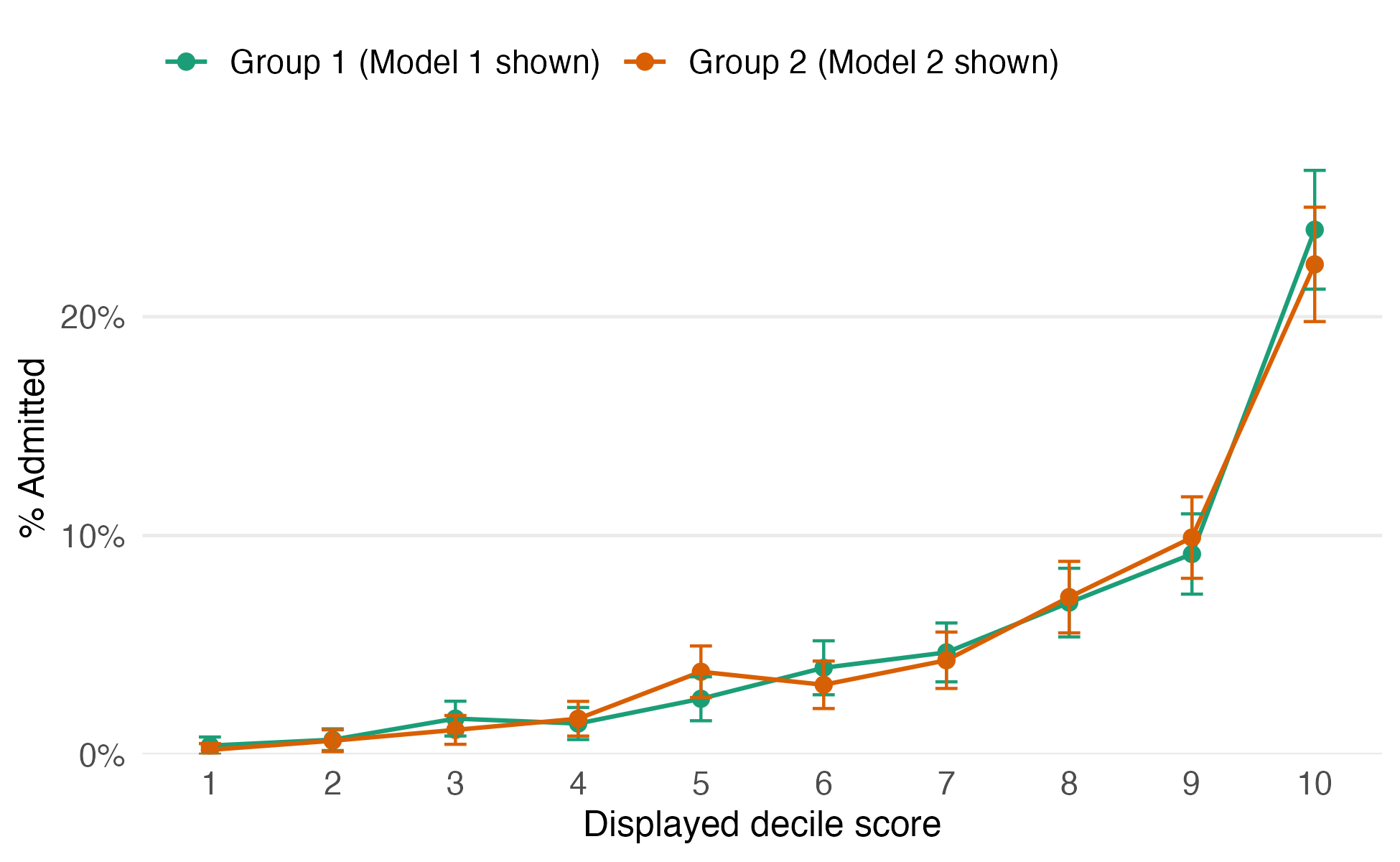}
\caption{Proportion of applicants admitted, by displayed algorithmic score decile, shown separately for applicants randomly assigned to have the Model~1 score (green) or Model~2 score (orange) displayed to admissions officers. Error bars indicate 95\% confidence intervals.}
\label{fig:2022}
\end{figure}

Figure~\ref{fig:2022} plots observed admission rates by presented score, shown separately for applicants for whom the Model~1 or Model~2 score was presented. We observe two things: first, in both conditions, admission rates increase monotonically with the presented score. Second, we observe that across the full score distribution, the two curves closely overlap. This indicates that the score scales produced by the two models are similarly calibrated with respect to observed decisions, despite the two models differing in their assessments of individual applicants.

Algorithmic scores are therefore consequential inputs, and the two models operate on comparable score scales. We next characterize the disagreement that the experimental manipulation introduces.

\subsection{What algorithmic disagreement does the experiment introduce?}
\label{sec:results-disagreement}

\subsubsection{How much do the models disagree at the individual level?}

We begin by quantifying the extent of individual-level disagreement between the two models. Although the models are similar in aggregate, they frequently diverge in their assessments of the same applicant. In our analytic sample, the two models assign different decile scores to 73.2\% of applicants. The mean absolute difference in assigned scores is 1.49 deciles (SD = 1.38), and the correlation between the two scores is moderate ($\rho = 0.75$).

\begin{figure*}[t]
  \centering
  \begin{minipage}{0.48\textwidth}
    \centering
    \includegraphics[width=\textwidth]{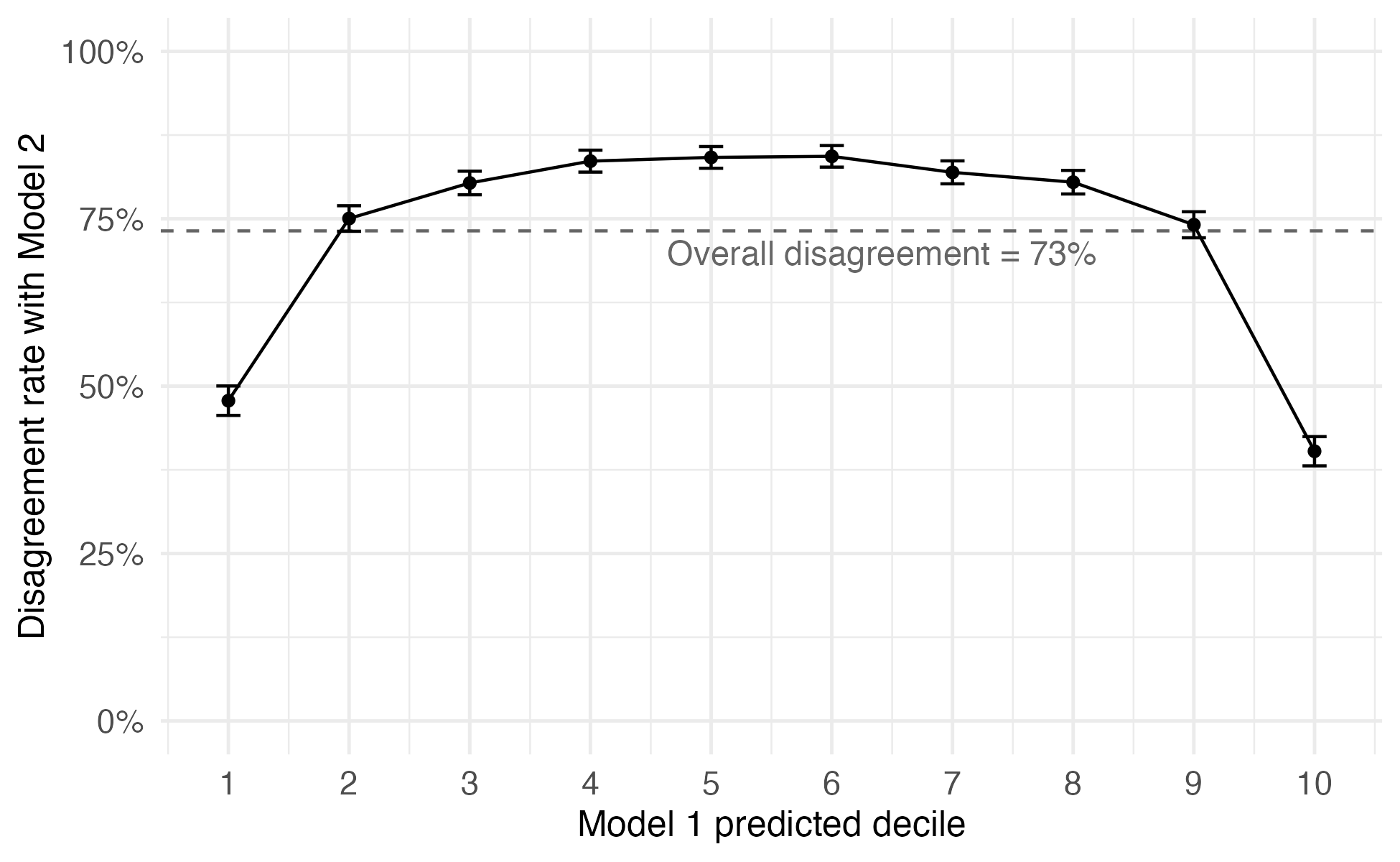}
    \caption*{(a) Disagreement rate by Model~1 decile}
  \end{minipage}
  \hfill
  \begin{minipage}{0.48\textwidth}
    \centering
    \includegraphics[width=\textwidth]{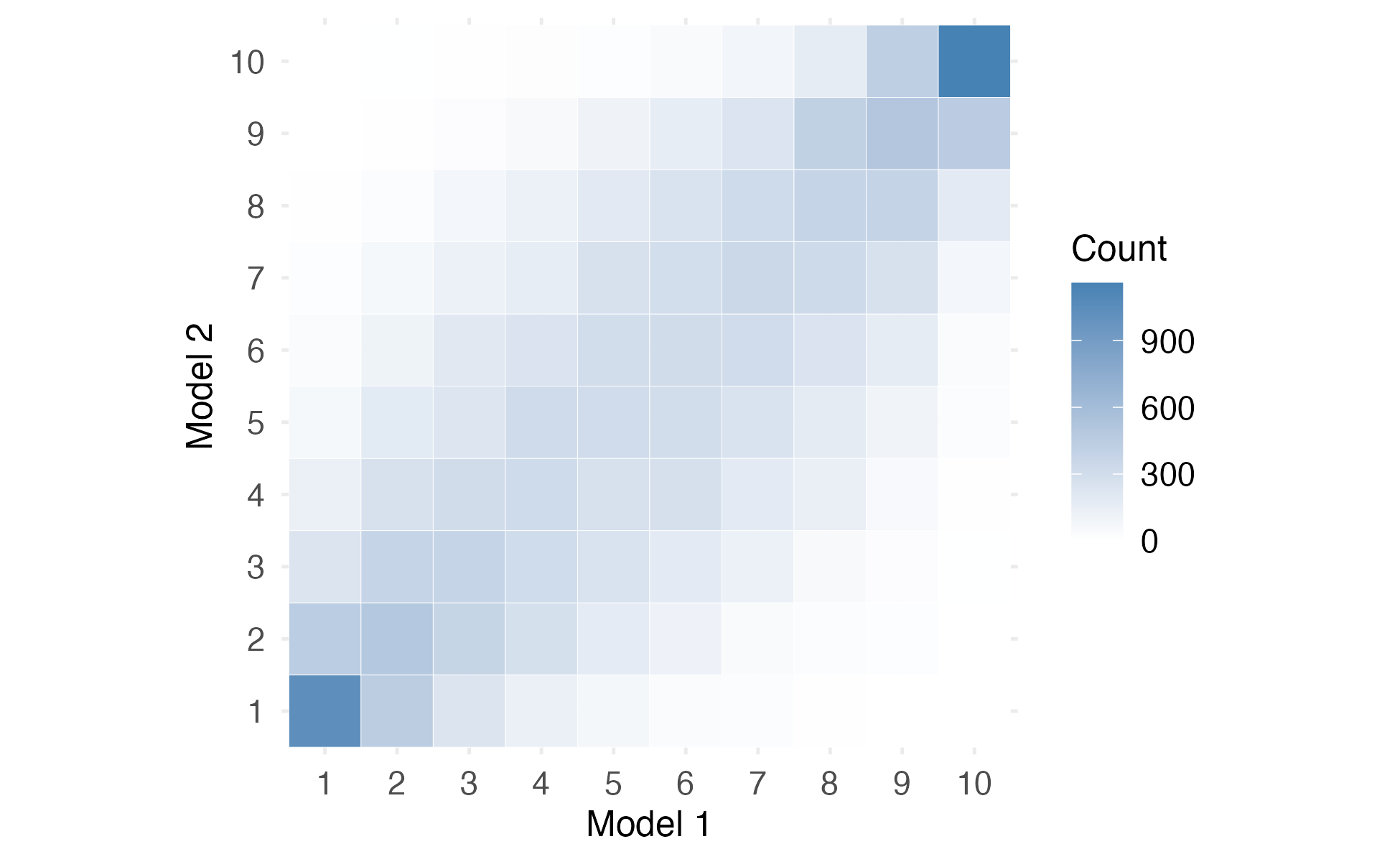}
    \caption*{(b) Joint distribution of Model~1 and Model~2 scores}
  \end{minipage}
  \caption{
  Individual-level disagreement between Model~1 and Model~2. Panel (a) shows the probability that the two models assign different decile scores, conditional on the Model~1 decile; the solid line connects per-decile disagreement rates with 95\% confidence interval error bars, and the dashed horizontal line denotes the overall disagreement rate (73\%) across all applicants. Panel (b) shows the joint distribution of decile scores assigned by the two models, where darker shading indicates a higher number of applicants and diagonal cells correspond to exact agreement.
  }
  \label{fig:model-disagreement}
\end{figure*}

Figure~\ref{fig:model-disagreement} illustrates both the prevalence and structure of this disagreement. Panel (a) shows that disagreement is lowest at the extremes of the score distribution, corresponding to clearer admit and clearer deny decisions, and highest in the middle of the score distribution, which contains the majority of applicants for whom admission decisions are less certain. Panel (b) shows that while many applicants lie near the diagonal, indicating agreement or near-agreement, there is substantial mass off the diagonal, including cases where the two models differ by three or more deciles.

Together, these patterns indicate that the experimental manipulation introduces meaningful variation in the algorithmic scores shown to admissions officers, concentrated among applicants for whom decision-making is most difficult. Because applicants are randomly assigned to which model’s score is shown, this disagreement creates exogenous variation in the algorithmic information presented to reviewers, providing the key source of variation we exploit in subsequent analyses to study algorithmic sensitivity.

\subsubsection{Is one model systematically more favorable?}

\begin{figure}[h!]
  \centering
  \includegraphics[width=0.55\columnwidth]{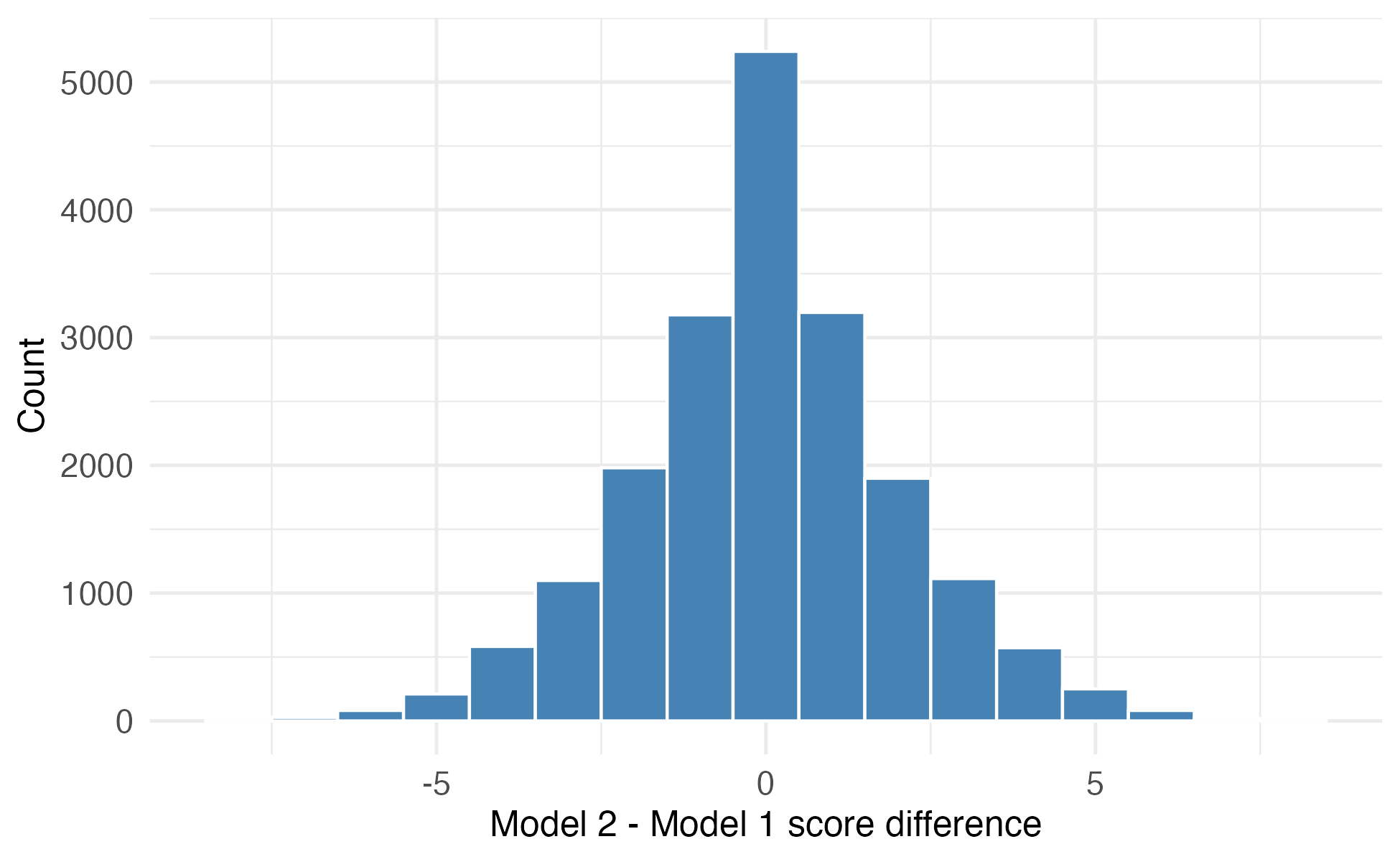}
  \caption{Distribution of signed score differences (Model~2 minus Model~1) on the analytic sample.}
  \label{fig:score_diff}
\end{figure}

The two models are symmetric in the aggregate. Mean predicted scores are nearly identical (5.4822 under Model~1, 5.4821 under Model~2), and the distribution of signed score differences is approximately symmetric and centered near zero (Figure~\ref{fig:score_diff}). Conditional on disagreement, the two directions occur with nearly equal frequency: $\omega_{1>2} = 0.501$ and $\omega_{2>1} = 0.499$ in the notation of Section~\ref{sec:estimating-ar}. Neither model is systematically more lenient, so the AS estimand does not mechanically favor one model's scores over the other; any estimated algorithmic sensitivity reflects responses to individual-level disagreement rather than global, systematic differences between the models. Aggregate symmetry masks modest subgroup-level variation (Model~2 assigns somewhat higher scores to URM and first-generation applicants), but as shown in Appendix~\ref{appendix:demographic}, these score-level differences do not translate into differential admission outcomes.

\subsection{Is there algorithmic sensitivity in admission decisions?}

\begin{table}[t]
\centering
\caption{Estimates of algorithmic sensitivity (effect of showing the higher vs.\ lower score)}
\label{tab:algorithmic-reliance}
\begin{tabular}{lcccccc}
\toprule
Case
& $N$
& $\mathbb{E}[Y \mid W{=}0]$
& $\mathbb{E}[Y \mid W{=}1]$
& $\widehat{\text{AS}}$
& SE
& $p$-value \\
\midrule

$S_{1} > S_{2}$ \; (Model~1 assigns the higher score)
& 7{,}166
& 0.0448
& 0.0420
& +0.0029
& 0.0048
& 0.554 \\

$S_{1} < S_{2}$ \; (Model~2 assigns the higher score)
& 7{,}139
& 0.0371
& 0.0427
& +0.0056
& 0.0046
& 0.225 \\

\midrule
$S_{1}\neq S_{2}$ \; (All disagreement cases, $\mathcal{D}$)
& 14{,}305
& ---
& ---
& +0.0042
& 0.0033
& 0.206 \\
\bottomrule
\end{tabular}

\begin{flushleft}
\footnotesize
\textit{Notes.}
$Y$ indicates admission.
$W=0$ denotes that the Model~1 score is shown and $W=1$ denotes that the Model~2 score is shown.
Rows labeled $S_1>S_2$ and $S_1<S_2$ report estimates of the stratum-specific components
$\text{AS}_{1>2}$ and $\text{AS}_{2>1}$ of algorithmic sensitivity as defined in Section~\ref{sec:estimating-ar}.
The pooled estimate $\widehat{\text{AS}}$ is the weighted average
$\omega_{1>2}\text{AS}_{1>2}+\omega_{2>1}\text{AS}_{2>1}$.
Positive values indicate that showing the more favorable score increases the probability of admission.
Standard errors are from linear probability models with HC3 robust standard errors.
\end{flushleft}
\end{table}

We now report estimates of the algorithmic sensitivity defined in Section~\ref{sec:estimating-ar}: the average causal effect, among applicants for whom the models disagree, of showing the more favorable rather than the less favorable algorithmic score. Of the 19{,}545 applicants in the analytic sample, the models disagree for 14{,}305 applicants (73.2\%). For the remaining 5{,}240 applicants (26.8\%), the models assign identical scores. This agreement group provides no identifying variation for algorithmic sensitivity,
because the favorability of the displayed score does not change with random assignment. Consistent with this, we observe no statistically detectable difference in admission outcomes across experimental arms (absolute difference $=0.34$ pp, SE $=0.79$ pp, $p=0.67$).

Table~\ref{tab:algorithmic-reliance} reports estimates of algorithmic sensitivity by disagreement direction. When Model~1 assigns the higher score ($S_1>S_2$), showing the more favorable rather than the less favorable score increases admission probability by an estimated 0.29 percentage points (SE = $0.48$ pp, $p=0.55$). When Model~2 assigns the higher score ($S_1<S_2$), showing the more favorable score increases admission probability by an estimated 0.56 percentage points (SE = 0.46 pp, $p=0.23$). Pooling across disagreement directions yields an overall estimate of $\widehat{\text{AS}}=0.42$ percentage points
(SE $=0.33$ pp, $p=0.21$). None of these estimates are statistically distinguishable from zero at conventional significance levels. This null result holds among higher-scoring applicants, for whom admission rates are substantially higher. Restricting to those receiving a score of 9 or 10 from at least one model yields a weighted average AS of $-0.005$, with only 2 strata reaching significance at the 5\% level in opposite directions (see Appendix~\ref{appendix:nine-ten}). AS estimates stratified by demographic group are similarly small and statistically indistinguishable from zero across all subgroups examined, including URM and first-generation applicants (Appendix~\ref{appendix:demographic}).

\subsection{Does algorithmic sensitivity vary with the magnitude of score disagreement?}
\begin{table*}[t]
\centering
\caption{Estimates of algorithmic sensitivity (effect of showing the higher vs.\ lower score) by magnitude of model disagreement}
\label{tab:algorithmic-reliance-by-gap}
\begin{tabular}{lcccccc}
\toprule
$|S_{1}-S_{2}|$
& $N$
& $\mathbb{E}[Y \mid W{=}0]$
& $\mathbb{E}[Y \mid W{=}1]$
& $\widehat{\text{AS}}$
& SE
& $p$-value \\
\midrule
\midrule
\multicolumn{7}{l}{\textit{Case 1: $S_{1} > S_{2}$ (Model~1 assigns the higher score)}} \\
\midrule
1 & 3{,}178 & 0.0442 & 0.0516 & $+0.0074$ & 0.0076 & 0.327 \\
2 & 1{,}981 & 0.0533 & 0.0426 & $-0.0108$ & 0.0096 & 0.262 \\
\textbf{3} & \textbf{1{,}099} & \textbf{0.0531} & \textbf{0.0157} & \textbf{-0.0374} & \textbf{0.0111} & \textbf{0.001} \\
4 & 583 & 0.0179 & 0.0461 & $+0.0281$ & 0.0144 & 0.052 \\
5 & 211 & 0.0094 & 0.0476 & $+0.0382$ & 0.0230 & 0.099 \\
6 & 82 & 0.0238 & 0.0000 & $-0.0238$ & 0.0238 & 0.326 \\
\midrule
Overall 
& 7{,}166
& 0.0448
& 0.0420
& +0.0029
& 0.0048
& 0.554 \\
\midrule
\midrule
\multicolumn{7}{l}{\textit{Case 2: $S_{1} < S_{2}$ (Model~2 assigns the higher score)}} \\
\midrule
1 & 3{,}197 & 0.0509 & 0.0480 & $-0.0029$ & 0.0077 & 0.705 \\
2 & 1{,}899 & 0.0307 & 0.0450 & $+0.0143$ & 0.0088 & 0.102 \\
3 & 1{,}114 & 0.0164 & 0.0265 & $+0.0102$ & 0.0087 & 0.242 \\
\textbf{4} & \textbf{573} & \textbf{0.0204} & \textbf{0.0573} & \textbf{+0.0369} & \textbf{0.0162} & \textbf{0.023} \\
\textbf{5} & \textbf{251} & \textbf{0.0472} & \textbf{0.0000} & \textbf{-0.0472} & \textbf{0.0190} & \textbf{0.013} \\
6 & 82 & 0.0244 & 0.0244 & $-0.0000$ & 0.0345 & 1.000 \\

\midrule
Overall
& 7{,}139
& 0.0371
& 0.0427
& +0.0056
& 0.0046
& 0.225 \\
\bottomrule
\end{tabular}

\begin{flushleft}
\footnotesize
\textit{Notes.}
$Y$ indicates admission.
$W=0$ denotes that the Model~1 score is shown and $W=1$ denotes that the Model~2 score is shown.
Within each panel, $\widehat{\text{AS}}$ denotes the estimated component of algorithmic sensitivity, where positive values indicate a higher probability of admission when the more favorable score is shown.
Rows labeled ``Overall'' correspond to $\widehat{\text{AS}}_{1>2}$ and $\widehat{\text{AS}}_{2>1}$ as reported in Table~\ref{tab:algorithmic-reliance}.
Standard errors are from linear probability models with HC3 robust standard errors.
\end{flushleft}
\end{table*}

Thus far, we have shown that algorithmic sensitivity is small and statistically indistinguishable from zero when aggregating across all disagreement cases. We next examine whether algorithmic sensitivity varies with the magnitude of disagreement between the two model scores. Specifically, we restrict attention to applicants in the disagreement set $\mathcal{D}$ and stratify the sample by the absolute score gap, $|S_{i1}-S_{i2}|$. Within each stratum and disagreement direction, estimated effects correspond to conditional versions of algorithmic sensitivity that hold fixed both the direction and magnitude of model disagreement.

Table~\ref{tab:algorithmic-reliance-by-gap} reports the resulting estimates. Across most values of $|S_{i1}-S_{i2}|$, estimated effects are small and statistically indistinguishable from zero. A small number of strata reach significance at the 5\% level, but the signs of these significant estimates are inconsistent across adjacent disagreement magnitudes within the same direction: $|S_{i1}-S_{i2}|=3$ in the $S_1>S_2$ case is negative and significant, while neighboring strata are positive; in the $S_1<S_2$ case, $|S_{i1}-S_{i2}|=4$ and $5$ are significant in opposite directions.

If decision-makers were increasingly influenced by score favorability as disagreement grew, we would expect effects to become larger in magnitude and more consistently signed at higher values of $|S_{i1}-S_{i2}|$. Instead, estimated effects fluctuate in both direction and size, suggesting that the isolated significant estimates are unlikely to reflect a consistent pattern of sensitivity.

\section{Discussion}
\label{sec:discussion}

We find little evidence that admission decisions track which of two equally defensible algorithmic scores reviewers see. Across the 14,305 applications for which the two models disagreed, often by several deciles and most often in the middle of the score distribution where decisions are least certain, estimated sensitivity effects are small, statistically indistinguishable from zero in pooled specifications, and inconsistent in direction across strata and subgroups. In this setting, the model-contingent variation that predictive multiplicity introduces does not propagate into final admission outcomes.

A null estimate of algorithmic sensitivity is, by itself, compatible with two distinct decision-making patterns. Reviewers may place little weight on the algorithmic score, so that variation in the displayed score does not move their decisions much. Alternatively, reviewers may place substantial weight on the score while discounting modest variation across models, on the implicit understanding that small differences in displayed score need not reflect substantively meaningful differences between applicants. Our design cannot distinguish between these patterns: both yield small sensitivity estimates, but they imply different things about how the algorithm functions in the decision process and about whether the absence of sensitivity here would generalize to settings in which the algorithm carries more weight in the decision.

This indeterminacy does, however, bound what mechanisms may be at work. If reviewers were strongly anchoring on the displayed score under time pressure, as participants do in laboratory settings \citep{rastogi2022deciding}, we would expect the more favorable score to produce systematically more favorable outcomes; we do not see this. If reviewers felt their decision-making authority was constrained by the algorithm and responded by deferring to whatever score was displayed, as decision-makers do in some algorithm-assisted settings \citep{jolly2026s}, we would similarly expect the score to drive outcomes; again, we do not. Whatever decision-making pattern is operative in our setting, it is not one in which the displayed score is propagated mechanically into outcomes.

If a mechanism of active discounting is at work, several features of the decision-making context plausibly contribute to it. Admissions decisions are holistic, integrating essays, recommendation letters, contextual background, and institutional priorities alongside the algorithmic score \cite{stevens2009creating}. The score is framed as a coarse, directional signal for attention allocation rather than as a recommendation, and is not used as a threshold at any stage of the process \cite{lee2023evaluating, lee2024algorithms}. Reviewers are experienced admissions professionals operating under institutional norms that legitimize discretion. The muted downstream influence we observe is consistent with prior field-based work showing that algorithmic input in real institutional settings is often more mediated by professional judgment than algorithm-centric accounts would predict \citep{imai2023experimental, stevenson2024algorithmic, brayne2021technologies}. Consistency with this broader pattern does not, however, establish that these specific features explain the attenuation we observe here, since our design varies the algorithmic input rather than the institutional context surrounding it.

Beyond the mechanism question, our finding has implications for how predictive multiplicity has been theorized as a problem. The literature has largely concentrated on properties of models: whether equally accurate models produce different predictions for the same individual, and what to do about it through ensembling, model selection, or calibration \citep{black2022model, marx2020predictive, cooper2024arbitrariness}. Our result addresses a different question, namely whether model-level multiplicity translates into outcome-level arbitrariness once the model is deployed in a human-mediated decision process. In our setting, it does not. \citet{du2025reconciling} make a complementary observation from the opposite direction: even when two predictive models agree on their individual predictions almost everywhere, they can lead downstream decision-makers to take substantially different actions. Taken together, these findings caution against assuming that what models do at the prediction level determines what decisions look like at the outcome level: the sociotechnical context in which a model is deployed can produce arbitrary outcomes from agreeing models or non-arbitrary outcomes from disagreeing ones. Which direction the context takes is itself an empirical question \citep{selbst2019fairness}, and our results provide one data point on the attenuating side.

Identifying these patterns required a framework that does not presume a notion of decision correctness, since the reliance literature's accuracy-based evaluation \citep{guo2024decision} is unavailable in many social decision settings, including admissions \citep{bruch2017decision}. The algorithmic sensitivity framework, by defining the estimand purely in terms of how decisions respond to changes in the algorithmic input, makes the propagation question tractable in such settings.

A null effect on algorithmic sensitivity is consistent with the absence of one specific harm---the systematic determination of individual outcomes by arbitrary modeling choices---but does not address the broader set of concerns motivating recent normative work on algorithmic decision-making. \citet{wang2024against} argue that predictive optimization in consequential domains is presumptively illegitimate because of a recurring set of flaws (target mismatch, distribution shift, limited contestability, susceptibility to gaming) that are not readily addressed by technical means; our results speak to one specific failure mode within this critique but do not address the broader argument. Models trained on historical admissions data may still encode patterns shaped by structural disadvantage, even when the choice between two such models does not produce differential outcomes for the groups we observe \citep{barocas2016big, gandara2024inside}. Algorithmic scoring may still obscure the basis for consequential decisions from applicants and the public \citep{burrell2016machine}. And the attenuation documented here should not be read as a general endorsement of human-in-the-loop arrangements: \citet{green2022flaws} shows that human oversight requirements have repeatedly been used to legitimize the deployment of algorithmic systems whose underlying problems remained unaddressed. Our results establish only that, in this setting, decisions were not systematically swayed by which of two model scores was shown; they do not establish that reviewers were performing the oversight that human-in-the-loop policy typically assumes, nor that the underlying models are appropriate to the decisions they inform. Although the experiment did not produce detectable disparate impact across the demographic groups we observed, individual applicants whose displayed score was less favorable than the alternative model would have produced bore a real consequence of the design.

\subsection{Limitations}

Four limitations bear on the interpretation of our findings. First, the variation we exploit comes from two specific models trained on overlapping but non-identical historical cohorts, rather than from randomized splits of identical data. Some of the disagreement between the two models therefore reflects cohort-level distributional shift \citep{quinonero2022dataset}, not just the pure sampling variability that most of the predictive multiplicity literature has focused on. This reflects realistic deployment conditions, since models in practice are retrained as new data arrive and any institution choosing between two such models encounters both sources of variation at once. But it means our estimates cannot be cleanly interpreted as sensitivity to sampling noise alone. Other forms of multiplicity, across model families, feature sets, or hyperparameter choices, may produce different patterns of disagreement and different patterns of sensitivity; whether our null extends to them is an open question.

Second, the algorithmic sensitivity estimand is identified on the subset of applicants for whom the two models disagree, which is not a random sample of the full applicant pool. Disagreement is concentrated in the middle score deciles, where admission rates are low (Figure~\ref{fig:model-disagreement}), so the pooled estimates speak most directly to applicants whose scores were not strongly signaling either admission or rejection. The robustness check in Appendix~\ref{appendix:nine-ten} addresses the case where sensitivity would matter most: applicants receiving high scores from at least one model, for whom admission rates are substantially higher and any sensitivity would have larger practical consequences. The null holds in this subgroup as well, but we cannot rule out that sensitivity exists in other subpopulations our analysis does not cover.

Third, our findings are specific to a setting with substantial procedural structure, experienced reviewers, and an algorithmic input framed as a coarse, directional signal rather than as a recommendation. We have argued that this configuration plausibly contributes to the attenuation we observe, but we have not varied it directly. Settings in which algorithmic scores are tightly coupled to decisions, in which reviewers lack expertise or discretion, or in which institutional pressure rewards mechanical compliance with algorithmic outputs may show very different patterns of sensitivity \citep{green2021algorithmic, stevenson2024algorithmic, wilson2025no}. Our design cannot identify which specific features of the institutional process are responsible for the attenuation, or whether the attenuation would survive in settings with different features.

Finally, our null result is bounded by statistical power. The 95\% confidence interval around the pooled estimate rules out effects larger than roughly one percentage point in either direction, but smaller effects are not detectable in our sample. Our data therefore rule out substantial algorithmic sensitivity in this setting, but they do not establish that decisions are exactly invariant to which score is shown.



\begin{acks}
This research was supported in part by NSF Awards IIS-2312865, OAC-2311521, and EDU-2237593. All content represents the opinion of the authors, which is not necessarily shared or endorsed by their respective employers and/or sponsors.
\end{acks}

\bibliographystyle{ACM-Reference-Format}
\bibliography{mybib}

\clearpage

\section*{Ethical Considerations Statement}

This study was conducted under a research protocol approved by an Institutional Review Board (IRB) and in close collaboration with senior admissions leadership at the participating institution. The research team worked alongside the institution's admissions office, which had independently determined that algorithmic scoring tools were needed to support holistic review under test-optional conditions. The widespread adoption of test-optional policies during the COVID-19 pandemic had eliminated standardized test scores as a universally available quantitative signal, while application volumes continued to rise, placing substantial time and attention constraints on admissions officers tasked with conducting individualized review of each application. Algorithmic scoring at the institution predated this study: a prior heuristic-based tool, which relied heavily on standardized test scores, had become ineffective under test-optional policies. Rather than having the admissions office develop new heuristics on their own, the research collaboration meant that the successor tools were developed and evaluated rigorously, and their effects on admissions decisions studied empirically, providing stronger empirical grounding than would typically be available in operational deployments without research involvement. All applicants still underwent human review, and the tools were not used to automate or replace human judgment. The admissions office was also an active partner in the research design: institutional leadership were specifically interested in understanding how sensitive their staff's decisions would be to algorithmic scores, a question with direct operational relevance to how these tools were deployed and communicated. All data were de-identified prior to analysis.

\paragraph{Potential for harm.} A central ethical concern in this design is whether presenting a lower score to some applicants constituted a disadvantage. We addressed this on two grounds prior to deployment. First, the two models were selected to be equally defensible: both achieved similar aggregate predictive performance and were trained on the same feature set and modeling pipeline. Although Model~2 performed slightly better on out-of-sample validation data, the inclusion of older training data increases the risk of distributional shift \cite{gama2014survey}, making it uncertain whether this advantage would persist in the upcoming cycle. Neither model was therefore clearly preferable, and no applicant was assigned a score from a model known to be inferior. Second, no applicant was automatically advanced or eliminated based on the score alone at any stage of the process; scores served exclusively as coarse, directional signals for attention allocation, with all final decisions made through holistic review by trained admissions professionals.

Our empirical findings further support these assurances. Appendix~\ref{appendix:demographic} reports predicted score distributions for both models by demographic group. Model~2 assigns modestly higher scores to URM and first-generation applicants on average (mean differences of 0.48 and 0.51 deciles respectively), likely reflecting year-to-year variation in admission patterns between the two training cohorts, while differences for other groups are negligible. These score-level differences did not translate into differences in admission outcomes: the two models were similarly calibrated with respect to observed admission decisions (Figure~\ref{fig:2022}), and AS estimates are small and statistically indistinguishable from zero across the main analysis (Table~\ref{tab:algorithmic-reliance}) and across demographic subgroups (Table~\ref{tab:ar-subgroup}). The randomization did not systematically disadvantage applicants in either condition. This does not establish that the models are free of bias in any broader sense, only that the randomization did not produce differential harm across the groups we observed.

\paragraph{Broader considerations.} Ongoing debates concern whether algorithmic tools should be used in university admissions decisions at all. Models trained on historical admissions data risk encoding patterns shaped by structural disadvantage: if prior admissions decisions reflected disadvantages facing certain groups, models trained on those decisions may systematically undervalue applicants from those groups \cite{barocas2016big, gandara2024inside}. Algorithmic scoring may also obscure the basis for consequential decisions from applicants and the public \cite{burrell2016machine}. Algorithmic input may subtly reshape holistic review by directing reviewer attention in ways that disadvantage already-marginalized applicants, even when no individual decision is explicitly automated. These concerns are not resolved by our findings. Our study addresses a specific, narrow question, namely whether human reviewers are sensitive to arbitrary variation in algorithmic scores, and a null result on this question is not an endorsement of algorithmic admissions tools broadly. In particular, our findings say nothing about whether the deployed models produce fair predictions, whether algorithmic scoring improves or undermines admissions decisions overall, or whether the use of such tools is appropriate in this context. The appropriate role of algorithms in high-stakes admissions decisions remains an open normative question for institutional leaders, affected communities, and researchers.


\appendix

\section{Causal Graph of the Experimental Design}
\label{appendix:graph}

\begin{figure}[h]
\centering
\begin{tikzpicture}[
  node distance    = 1.8cm and 2.6cm,
  box/.style       = {draw, rectangle, thick,
                      minimum width=2.8cm, minimum height=0.8cm,
                      align=center, font=\small\upshape},
  exo/.style       = {draw, rectangle, thick, double, double distance=2pt,
                      minimum width=2.8cm, minimum height=0.8cm,
                      align=center, font=\small\upshape},
  dim/.style       = {draw, rectangle, thick, dashed, text opacity=0.3,
                      draw opacity=0.3,
                      minimum width=2.8cm, minimum height=0.8cm,
                      align=center, font=\small\upshape},
  arr/.style       = {-{Stealth[length=5pt]}, thick},
  arr_focal/.style = {-{Stealth[length=5pt]}, very thick},
  arr_rnd/.style   = {-{Stealth[length=5pt]}, thick, dashed},
  arr_dim/.style   = {-{Stealth[length=5pt]}, thick, opacity=0.3}
]

\node[box]                              (X)  {Application\\materials};

\node[box, below left=1.8cm and 2.0cm of X]   (S1) {Model 1\\score};
\node[box, below right=1.8cm and 2.0cm of X]  (S2) {Model 2\\score};

\node[box, below=3.8cm of X]                  (Ss) {Displayed\\score};
\node[exo, left=2.6cm of Ss]                  (W)  {Random\\assignment};

\node[box, below=1.8cm of Ss]                 (H)  {Human\\review};

\node[box, below=1.8cm of H]                  (Y)  {Admission\\decision};
\node[dim, right=2.6cm of Y]                  (O)  {Downstream\\outcomes};

\draw[arr]       (X)  -- (S1);
\draw[arr]       (X)  -- (S2);
\draw[arr]       (X)  to[bend left=40] (H);
\draw[arr]       (S1) -- (Ss);
\draw[arr]       (S2) -- (Ss);
\draw[arr_rnd]   (W)  -- (Ss);
\draw[arr_focal] (Ss) -- (H);
\draw[arr_focal] (H)  -- (Y);
\draw[arr_dim]   (Y)  -- (O);

\end{tikzpicture}
\caption{Causal graph of the experimental design. The bold path from displayed score to human review to admission decision 
represents the AS estimand: the causal effect of the displayed algorithmic score on the admission decision, mediated through 
the human reviewer's judgment. Random assignment (double border) generates exogenous variation in the displayed score. The dimmed node and edge indicate that downstream decision outcomes are outside the scope of this study.}
\label{fig:dag}
\end{figure}

\section{Model Score Distributions and Algorithmic Sensitivity by Demographic Group}
\label{appendix:demographic}

\begin{figure}[ht]
\centering
\includegraphics[width=\textwidth]{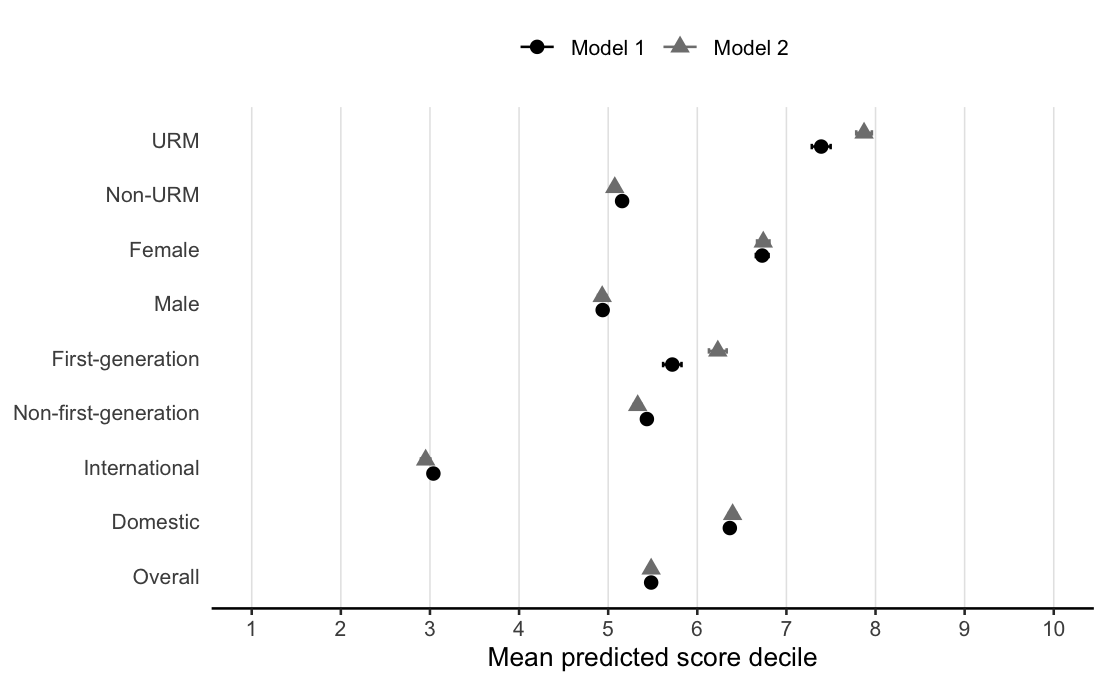}
\caption{Mean predicted score decile by demographic group for Model~1 and Model~2. Error bars indicate 95\% confidence 
intervals. The two models are closely aligned for most groups; Model~2 assigns modestly higher scores to URM and first-generation applicants.}
\label{fig:score-demo}
\end{figure}

Figure~\ref{fig:score-demo} shows mean predicted score deciles for Model~1 and Model~2 by demographic group, with 95\% confidence intervals. The two models produce closely aligned score distributions for most groups. Model~2 assigns modestly higher scores to URM applicants (mean difference = 0.48 deciles) and first-generation applicants (mean difference = 0.51 deciles) relative to Model~1, while differences for other groups are negligible. Table~\ref{tab:score-demo} reports the corresponding summary statistics.

These subgroup-level differences likely reflect year-to-year variation in admission patterns across the two training cohorts: because URM and first-generation applicants are evaluated more holistically and their admission rates are more sensitive to institutional priorities that may shift across cycles, small differences in training data composition disproportionately affect their predicted scores relative to groups whose admission patterns are more stable \cite{quinonero2022dataset}.

Despite these score-level differences, Table~\ref{tab:ar-subgroup} shows that AS estimates are small and statistically indistinguishable from zero across all demographic groups, indicating that the modest score differences between models did not propagate into differential admission outcomes for any group. Taken together, these results suggest that neither model systematically disadvantaged any demographic subgroup in terms of realized admission decisions.

\begin{table}[ht]
\centering
\caption{Mean predicted score decile by demographic group.}
\label{tab:score-demo}
\begin{tabular}{lcccccc}
\toprule
Group & $N$ & 
  $\bar{S}_1$ & $\bar{S}_2$ & 
  Mean diff. & SD diff. \\
\midrule
Overall              & 19,545 & 5.482 & 5.482 & $+0.000$ & 2.031 \\
\midrule
URM                  &  2,852 & 7.391 & 7.871 & $+0.481$ & 1.808 \\
Non-URM              & 16,693 & 5.156 & 5.074 & $-0.082$ & 2.055 \\
\midrule
Female               &  5,940 & 6.728 & 6.740 & $+0.012$ & 2.004 \\
Male                 & 13,605 & 4.938 & 4.933 & $-0.005$ & 2.042 \\
\midrule
First-generation     &  3,285 & 5.720 & 6.230 & $+0.511$ & 1.968 \\
Non-first-generation & 16,260 & 5.434 & 5.331 & $-0.103$ & 2.028 \\
\midrule
International        &  5,187 & 3.038 & 2.952 & $-0.087$ & 1.907 \\
Domestic             & 14,358 & 6.365 & 6.396 & $+0.031$ & 2.073 \\
\bottomrule
\end{tabular}
\begin{flushleft}
\footnotesize
\textit{Notes.} $\bar{S}_1$ and $\bar{S}_2$ denote mean predicted 
score deciles for Model~1 and Model~2 respectively. Mean diff.\ 
is Model~2 minus Model~1. SD diff.\ is the standard deviation 
of the within-applicant score difference.
\end{flushleft}
\end{table}

\begin{table}[ht]
\centering
\caption{Estimates of algorithmic sensitivity by demographic group.}
\label{tab:ar-subgroup}
\begin{tabular}{lcccc}
\toprule
Group & $N$ & $\widehat{\text{AS}}$ & SE & $p$-value \\
\midrule
URM                  &  1,740 & $+0.0156$ & 0.0124 & 0.210 \\
Non-URM              & 12,565 & $+0.0025$ & 0.0034 & 0.467 \\
\midrule
Female               &  4,210 & $+0.0002$ & 0.0072 & 0.983 \\
Male                 & 10,095 & $+0.0057$ & 0.0036 & 0.118 \\
\midrule
First-generation     &  2,342 & $+0.0052$ & 0.0088 & 0.552 \\
Non-first-generation & 11,963 & $+0.0040$ & 0.0036 & 0.264 \\
\midrule
International        &  3,540 & $+0.0013$ & 0.0032 & 0.680 \\
Domestic             & 10,765 & $+0.0049$ & 0.0043 & 0.254 \\
\bottomrule
\end{tabular}
\begin{flushleft}
\footnotesize
\textit{Notes.} Sample restricted to applicants for whom the two 
models disagree. $\widehat{\text{AS}}$ is the estimated effect of 
being shown the more favorable score on admission probability, 
estimated via linear probability model with HC3 robust standard 
errors. No estimate is statistically significant at the 5\% level.
\end{flushleft}
\end{table}

\section{Algorithmic Sensitivity Among Higher-Scoring Applicants}
\label{appendix:nine-ten}

A potential concern with the main analysis is that disagreement cases are concentrated in middle deciles where admission rates are low, such that near-zero estimates of algorithmic sensitivity could partly reflect a floor effect rather than genuine invariance. To address this, we restrict attention to applicants receiving a score of 9 or 10 from at least one model and for whom the two models disagree ($n = 3{,}561$), where admission rates are substantially higher and algorithmic sensitivity would be most detectable if present.

Table~\ref{tab:algorithmic-reliance-by-gap-highscores} reports estimates stratified by the magnitude of model disagreement within 
this subset. The weighted average AS is $-0.005$. Only 2 strata reach significance at the 5\% level, and these are in opposite directions, providing no evidence of a consistent pattern of sensitivity. The null result thus holds precisely where sensitivity would be most detectable, alleviating concerns that the main findings are an artifact of floor effects among lower-scoring applicants.

\begin{table*}[t]
\centering
\caption{Estimates of algorithmic sensitivity by magnitude of model 
disagreement (subset: applicants scored 9 or 10 by at least one 
model)}
\label{tab:algorithmic-reliance-by-gap-highscores}
\begin{tabular}{ccccccc}
\toprule
$|S_{1}-S_{2}|$
& $N$
& $\mathbb{E}[Y \mid W{=}0]$
& $\mathbb{E}[Y \mid W{=}1]$
& $\widehat{\text{AS}}$
& SE
& $p$-value \\
\midrule
\multicolumn{7}{l}{\textit{Case 1: $S_{1} > S_{2}$ 
  (Model~1 assigns the higher score)}} \\
1 & 838 & 0.1019 & 0.1202 & $+0.0183$ & 0.0218 & 0.401 \\
2 & 449 & 0.1132 & 0.0928 & $-0.0204$ & 0.0289 & 0.481 \\
\textbf{3} & \textbf{244} & \textbf{0.0957} & \textbf{0.0310}
  & $\mathbf{-0.0646}$ & \textbf{0.0317} & \textbf{0.042} \\
4 & 134 & 0.0345 & 0.0658 & $+0.0313$ & 0.0377 & 0.408 \\
5 & 75  & 0.0000 & 0.0227 & $+0.0227$ & 0.0230 & 0.326 \\
6 & 31  & 0.0000 & 0.0000 & $+0.0000$ & 0.0000 & --- \\
\midrule
\multicolumn{7}{l}{\textit{Case 2: $S_{1} < S_{2}$ 
  (Model~2 assigns the higher score)}} \\
1 & 841 & 0.1394 & 0.1181 & $-0.0213$ & 0.0232 & 0.358 \\
2 & 405 & 0.0686 & 0.1244 & $+0.0558$ & 0.0294 & 0.059 \\
3 & 242 & 0.0268 & 0.0308 & $+0.0040$ & 0.0217 & 0.854 \\
\textbf{4} & \textbf{149} & \textbf{0.0405} & \textbf{0.1333}
  & $\mathbf{+0.0928}$ & \textbf{0.0461} & \textbf{0.046} \\
5 & 75  & 0.0811 & 0.0000 & $-0.0811$ & 0.0461 & 0.083 \\
6 & 37  & 0.0000 & 0.0000 & $+0.0000$ & 0.0000 & --- \\
\bottomrule
\end{tabular}
\begin{flushleft}
\footnotesize
\textit{Notes.}
$Y$ indicates admission.
$W=0$ denotes that the Model~1 score is shown and $W=1$ denotes 
that the Model~2 score is shown.
Within each panel, $\widehat{\text{AS}}$ denotes the estimated 
component of algorithmic sensitivity, where positive values indicate 
a higher probability of admission when the more favorable score 
is shown. Boldface rows indicate estimates significant at the 
5\% level.
Standard errors are from linear probability models with HC3 
robust standard errors.
\end{flushleft}
\end{table*}

\section*{Generative AI Usage Statement}
Generative AI tools were used in a limited capacity during manuscript preparation to assist with grammar and fluency editing of author-written text, as well as with formatting tables. Generative AI tools were not used to generate substantive text, arguments, results, analyses, or interpretations. All intellectual contributions, empirical analyses, and conclusions are solely those of the authors, who take full responsibility for the originality, accuracy, and integrity of the manuscript.

\end{document}